\documentclass[runningheads,a4paper]{llncs}
\usepackage[T1]{fontenc}
\usepackage{amsfonts}
\usepackage{amsmath}
\usepackage{amssymb}
\usepackage{amsbsy}
\usepackage{mathtools}
\usepackage{color}
\usepackage{xcolor}
\usepackage{graphicx}
\usepackage{array}
\usepackage{enumerate}
\usepackage{enumitem}
\usepackage{sectsty}
\usepackage{appendix}
\usepackage{multirow}
\usepackage{cite}
\usepackage{tensor}
\usepackage{algorithm}
\usepackage{algorithmicx}
\usepackage{algpseudocode}
\usepackage{listings}
\usepackage{ulem} 
\usepackage{bbold} 
\usepackage{setspace}
\usepackage{epstopdf}
\usepackage{epsfig}
\usepackage{pgf}
\usepackage{tikz}
\usetikzlibrary{shapes,snakes,arrows}
\usepackage{subfigure}
\usepackage{url}
\urldef{\mailsa}\path|lichtenstein@physik.rwth-aachen.de|
\urldef{\mailsb}\path|e.di.napoli@fz-juelich.de|
\urldef{\mailsc}\path|dinapoli@aices.rwth-aachen.de|  
\newcommand{\keywords}[1]{\par\addvspace\baselineskip
\noindent\keywordname\enspace\ignorespaces#1}

\newcommand\be{\begin{equation}}
\newcommand\ee{\end{equation}}
\newcommand\bea{\begin{eqnarray}}
\newcommand\eea{\end{eqnarray}}
\newcommand\bi{\begin{itemize}}
\newcommand\ei{\end{itemize}}

\newcommand{\err}[1]{\ensuremath{\textsc{err}[#1]}}
\newcommand{\tus}{\textunderscore}
\newcommand{\Chi}{\mathcal{X}}

\newcommand\eqalign[1]{%
	\vcenter{%
		\normalbaselines \advance\baselineskip 5pt
		\advance\lineskip 5pt \tabskip=0pt
		\halign{%
			&\hfil $\displaystyle{##{}}$&
			$\displaystyle{{}##}$\hfil\cr
			#1\crcr
			}%
		}%
	}

\newcommand\dotline{\par\hbox to \hsize{\dotfill}\par}

\makeatletter
\def\befored@t#1.#2.#3;{#1}
\def\afterd@t#1.#2.#3;{#2}
\def\refhead#1{\edef\next{\ref{#1}}\expandafter\befored@t\next..;}
\def\reftail#1{\edef\next{\ref{#1}}\expandafter\afterd@t\next..;}
\def\lsim{\mathrel{\mathpalette\@versim<}}
\def\gsim{\mathrel{\mathpalette\@versim>}}
\def\@versim#1#2{\vcenter{\offinterlineskip
        \ialign{$\m@th#1\hfil##\hfil$\crcr#2\crcr\sim\crcr } }}
\newcommand\becomes[1]{\mathchoice{\becomes@\scriptstyle{#1}}
   {\becomes@\scriptstyle{#1}} {\becomes@\scriptscriptstyle{#1}}
   {\becomes@\scriptscriptstyle{#1}}}
\def\becomes@#1#2{\mathrel{\setbox0=\hbox{$\m@th #1{\,#2\,}$}%
        \mathop{\hbox to \wd0 {\rightarrowfill}}\limits_{#2}}}
\makeatother

\newcommand\pluseqq{\mathrel{+}=}
\newcommand\minuseqq{\mathrel{-}=}

\algnewcommand{\IIf}[1]{\State\algorithmicif\ #1\ \algorithmicthen}
\algnewcommand{\EndIIf}{\unskip\ \algorithmicend\ \algorithmicif}

\colorlet{graybg}{gray!10}
\lstdefinelanguage{alg}{
    keywords=[1]{for,try,success,failure,if,then,else,while,do,return},
    keywordstyle=[1]{\bf},
}
\begin{document}
%
\mainmatter

\title{Parallel adaptive integration in high-performance functional
  Renormalization Group computations}

\titlerunning{Adapative integration in HPC fRG}
            
\author{Julian Lichtenstein\inst{1}
\and Jan Winkelmann\inst{2} \and David S\'anchez de la Pe\~na\inst{1}
\and Toni Vidovi\'c\inst{3} \and Edoardo Di Napoli\inst{2,4}}%

\institute{Institute for Theoretical Solid State Physics, RWTH
  Aachen University.\\
\and
Aachen Institute for Advance Study in Computational Engineering Science.\\
\and
Department of Mathematics, Faculty of Science, University of Zagreb.\\
\and J\"ulich Supercomputing Centre, 
Forschungszentrum J\"ulich.
}

\authorrunning{Lichtenstein et al.}

\toctitle{Lecture Notes in Computer Science}
\tocauthor{Julian Lichtenstein}
\maketitle

\begin{abstract} 
  The conceptual framework provided by the functional Renormalization
  Group (fRG) has become a formidable tool to study correlated
    electron systems on lattices which, in turn, provided great
    insights to our understanding of complex many-body phenomena, such
    as high-temperature superconductivity or topological states of
    matter. In
  this work we present one of the latest realizations of fRG which
  makes use of an adaptive numerical quadrature scheme specifically tailored
  to the described fRG scheme. The final result is an increase in
  performance thanks to improved parallelism and scalability.

\end{abstract}


\section{Introduction} 
\label{sec:intro}


In this paper we report on the algorithmic and performance
improvements resulting from the collaboration between High-Performance
Computing (HPC) experts and domain scientists in the specific field of
functional Renormalization Group (fRG). In particular, we focus on an
adaptive implementation of a two-dimensional numerical quadrature
algorithm tailored to the evaluation of a large number of integrals
within a recently developed fRG method. The result of such an effort
is the Parallel Adaptive Integration in two Dimensions (PAID)
library. PAID requires approximately an order of magnitude fewer
operations for the computation of the numerical integrals and
translates this reduction into a substantial gain in parallel
performance. 

The Renormalization Group (RG) is a powerful method describing the
behavior of a physical system at different energy
and length scales. RG techniques allow smooth interpolation between
well studied models at a given energy scale and complicated emergent
phenomena at lower energy scales. In what is known as {\it RG flow},
physical quantities are computed iteratively with respect to variation
of an auxiliary scale parameter by solving a system of coupled
ordinary integro-differential equations.  In its application to
interacting electrons on a lattice at low temperatures, fRG methods
are commonly used to detect transitions of the metallic state towards
some ordered state\cite{Metzner2012,Platt2013}.  At the initial energy
scale of the flow, the physical system is in a well understood
metallic state, where weakly correlated electrons interact pairwise
through Coulomb repulsion. Lowering the scale, a second order phase
transition may take place at some critical temperature, in which some
form of order (e.g. magnetism or superconductivity) spontaneously
emerges.

fRG methods passed through many refinements over the
years\cite{Zanchi1998,Salmhofer2001,Husemann2009,Wang2012} in the form
of specific approximation schemes.  While each of these schemes has
its strength, the improvement of their accuracy (e.g. on predictions
for critical temperatures) is still underway.
%
%
In the current work, we illustrate the \textit{Truncated Unity} scheme
(TUfRG)~\cite{Lichtenstein2016} and its parallel implementation. One
of the computational advantages of this scheme stems from the
insertion of truncated partitions of unity in the flow equations. The
resulting numerical integration becomes less challenging at the
expense of having extra operations to perform (so-called inter-channel
projections). At each step of the
equations' flow one ends up computing multiple independent integrals
parametrized by three indices, namely $\mathbf{l}$, $m$, and $n$. In the
original TUfRG code all these integrals are distributed over a large
number of threads where each one is computed sequentially by a single
thread using the adaptive DCUHRE library~\cite{Berntsen1991}. Since
computing such integrals accounts for around 80\% of the total
computational time, they are the ideal candidate for an HPC
optimization.

While recent implementations have shown increased performance and
scalability through parallel quadrature
schemes~\cite{DApuzzo:1997ud,Laccetti:1999tv}, we follow a different
path by tailoring the numerical quadrature to the needs of the TUfRG
algorithm. In PAID, the subset of all integrals corresponding to one
value of $l$ are collected in a container and computed adaptively. All
the integrals in the container become tasks which are executed under
just one parallel region over the shared memory of a compute
node. With this approach we intend to gain better control over the
global quadrature error and minimize load imbalance while increasing
scalability.  Our results show that PAID scales as well as the trivial
parallelization using DCUHRE. In addition, PAID's adaptivity over the indexes
$m$ and $n$ of the integrals consistently yields a speedup from
$2\times$ up to $4\times$. In section 2 we give a brief account of
the method at the base of the TUfRG scheme. In the following section
we present the basic notion of adaptive integration and the algorithm
underlying the PAID library. In section 4 we describe the parallel
implementation in more detail. We conclude with a section on numerical
results and future work.



\section{The fRG method and the Truncated Unity approach}
\label{sec:tufrg}

In this section we provide a short introduction to the mathematical
framework of the TUfRG scheme. This is by no means an exhaustive
description and we refer the reader to~\cite{Lichtenstein2016} for a
detailed presentation. As several other fRG methods, TUfRG focuses on
interacting electrons on 2D lattice systems. These systems exhibit
strong correlations at low energy, which results in 
a rich diversity of ordered ground states. At the mathematical level, the effective
two-particle coupling function
$V(k_{0,1},\mathbf{k_1};k_{0,2},\mathbf{k_2};k_{0,3},\mathbf{k_3})$
contains essential information on the properties of the electronic
ground state. $V$ depends on both three frequencies ($k_0$) and three
momenta ($\mathbf{k}$), while the fourth ones are fixed due to
conservation of energy and momentum respectively. In favor of a short
notation, we sum up the dependence on frequency and momentum of one
particle into a combined index $k=(k_0,\mathbf{k})$ and write
$V(k_1,k_2,k_3)$.

The fRG calculation is based on the insertion of a control parameter
$\Omega$, which is an artificial energy scale. It is used for tuning
the system from an easily solvable starting point to a system that
includes the physically important features. For a given value of the control
parameter, the strong correlations at energies below $\Omega$ 
are excluded from $V$.
Starting from a high enough $\Omega$
results in a well-defined initial value for $V$, which corresponds to an 
interaction between two isolated charges. By decreasing $\Omega$, we
successively include correlation effects into the effective
two-particle coupling. From a mathematical point of view, the
calculation of $V$ at lower energy scales can be seen as an {\it
  initial value problem}, where the value of $V$ at a high energy
scale is the initial value. In order to obtain the resulting
two-particle coupling function at lower values of $\Omega$, one needs
to integrate a first order ordinary differential equation extracted
from a so-called level-2 truncation of the fRG equation
hierarchy\cite{Metzner2012} and from neglecting self-energies. Such an
equation can be written as
\begin{equation}
    \label{eqn:su2fl}
 \dot{V} (k_1,k_2,k_3) = \mathcal{T}_\mathrm{pp} (k_1,k_2,k_3)
 +  \mathcal{T}^\mathrm{cr}_\mathrm{ph}  (k_1,k_2,k_3) + \mathcal{T}^\mathrm{d}_\mathrm{ph} (k_1,k_2,k_3) \, ,
\end{equation}
where the dependence on $\Omega$ of all quantities is implicit and the
dot above $V$ denotes the first derivative with respect to the
artificial energy scale. The right-hand side is divided in three main
contributions: a particle-particle
\begin{equation} \label{eqn:pp-su2fl}
 \mathcal{T}_\mathrm{pp} = - \int \! d p \, \left[ \partial_\Omega G(p) \, G(k_1+k_2-p) \right]
  V (k_1,k_2,p) \, V(k_1+k_2-p,p,k_3) \, ,
\end{equation}
a crossed particle-hole 
\begin{equation} \label{eqn:phcr-su2fl}
 \mathcal{T}^\mathrm{cr}_\mathrm{ph} = - \int \! d p \, \left[ \partial_\Omega G(p) \, G(p+k_3-k_1) \right]
 V (k_1,p+k_3-k_1,k_3) \, V(p,k_2,p+k_3-k_1)
\end{equation}
and three direct particle-hole terms summarized in
$\mathcal{T}^\mathrm{d}_\mathrm{ph}$ as
\begin{align} \notag
 \mathcal{T}^\mathrm{d}_\mathrm{ph} =  \int \! d p & \,\, \left[ \partial_\Omega G(p) \, G(p+k_2-k_3) \right] 
 \left[ 2 V (k_1,p+k_2-k_3,p) \, V(p,k_2,k_3) \right. \\  \notag
&  -  V (k_1,p+k_2-k_3,k_1+k_2-k_3) \, V(p,k_2,k_3)   \\ \label{eqn:phd-su2fl} & \left.
 -  V (k_1,p+k_2-k_3,p) \, V(p,k_2,p+k_2-k_3) \right] \, .
\end{align}
All five summands that appear as integrands are quadratic in both $V$ and the 
function $G(k)=\frac{\theta(k)}{ik_0-\epsilon(\mathbf{k})}$, which is 
the propagator of the system containing non-interacting particles. 
The regulator function $\theta$ implements the exclusion of 
correlation effects from the system at energies below $\Omega$. In this 
paper we use $\theta(k)=\theta(k_0)=\frac{k_0^2}{k_0^2+\Omega^2}$ as
regulator, which suppresses $G$ for $\Omega$ much larger than all
relevant energy scales of the system. In the limit of $\Omega \to 0$
the structure of $G$ is recovered and we regain the physical 
system. The energy dispersion $\epsilon(\mathbf{k})$---which appears 
in the denominator of $G$---contains the energy spectrum of the 
single-particle problem. Since in this paper we are dealing with a 
$t$-$t'$ Hubbard model on a square lattice, the dispersion is
\begin{equation}
    \epsilon(\mathbf{k})=-2\,t\,(\cos(k_x)+\cos(k_y))-4\,t'\,\cos(k_x)\,\cos(k_y)-\mu
\end{equation}
where $t$ and $t'$ describe the kinetics of the particles and $\mu$ 
is the chemical potential controlling the total number of particles 
in the system.

\begin{figure}[t]
    \centering
    \includegraphics[width=0.45\textwidth]{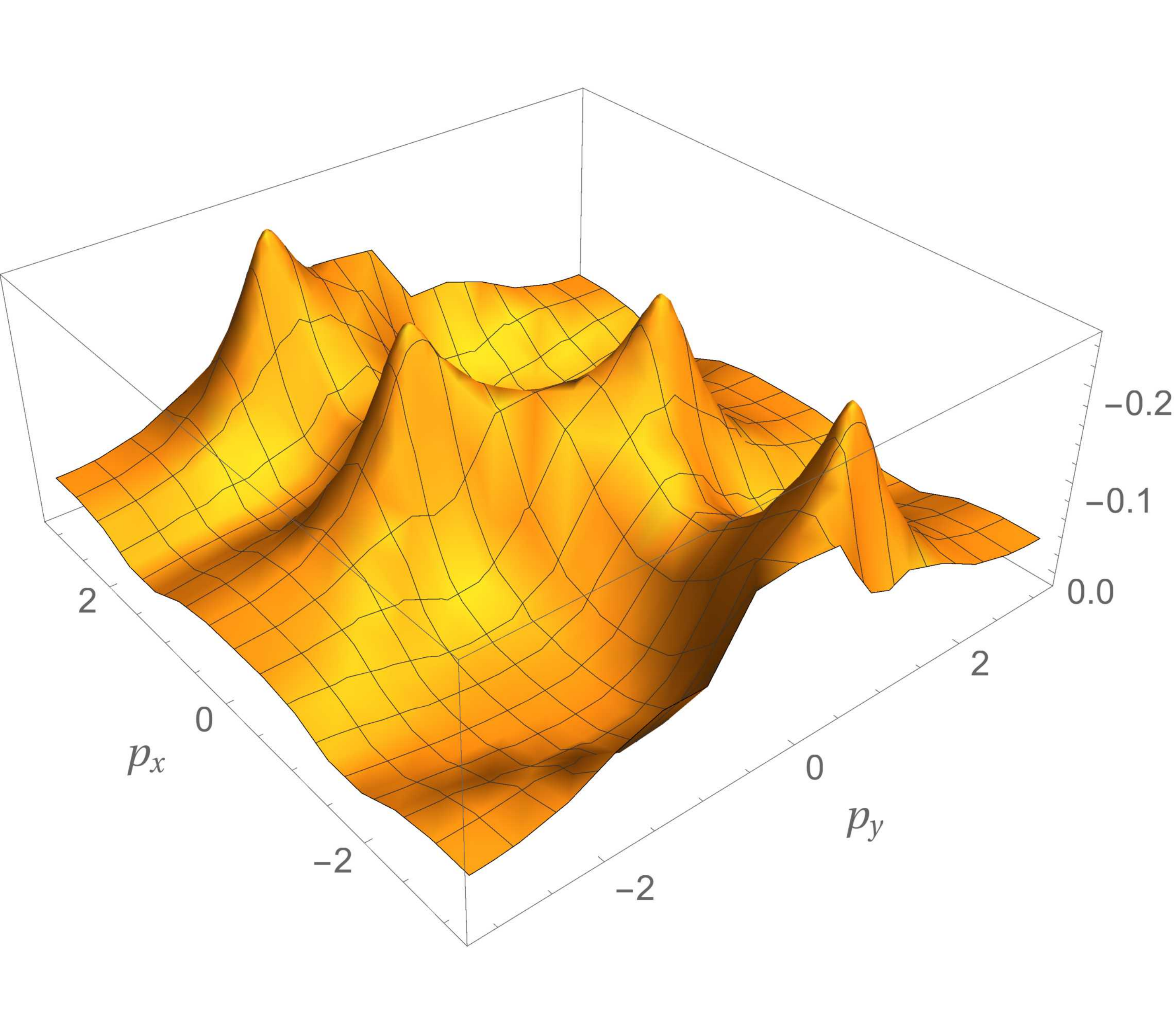}
    \hspace{0.08\textwidth}
    \includegraphics[width=0.45\textwidth]{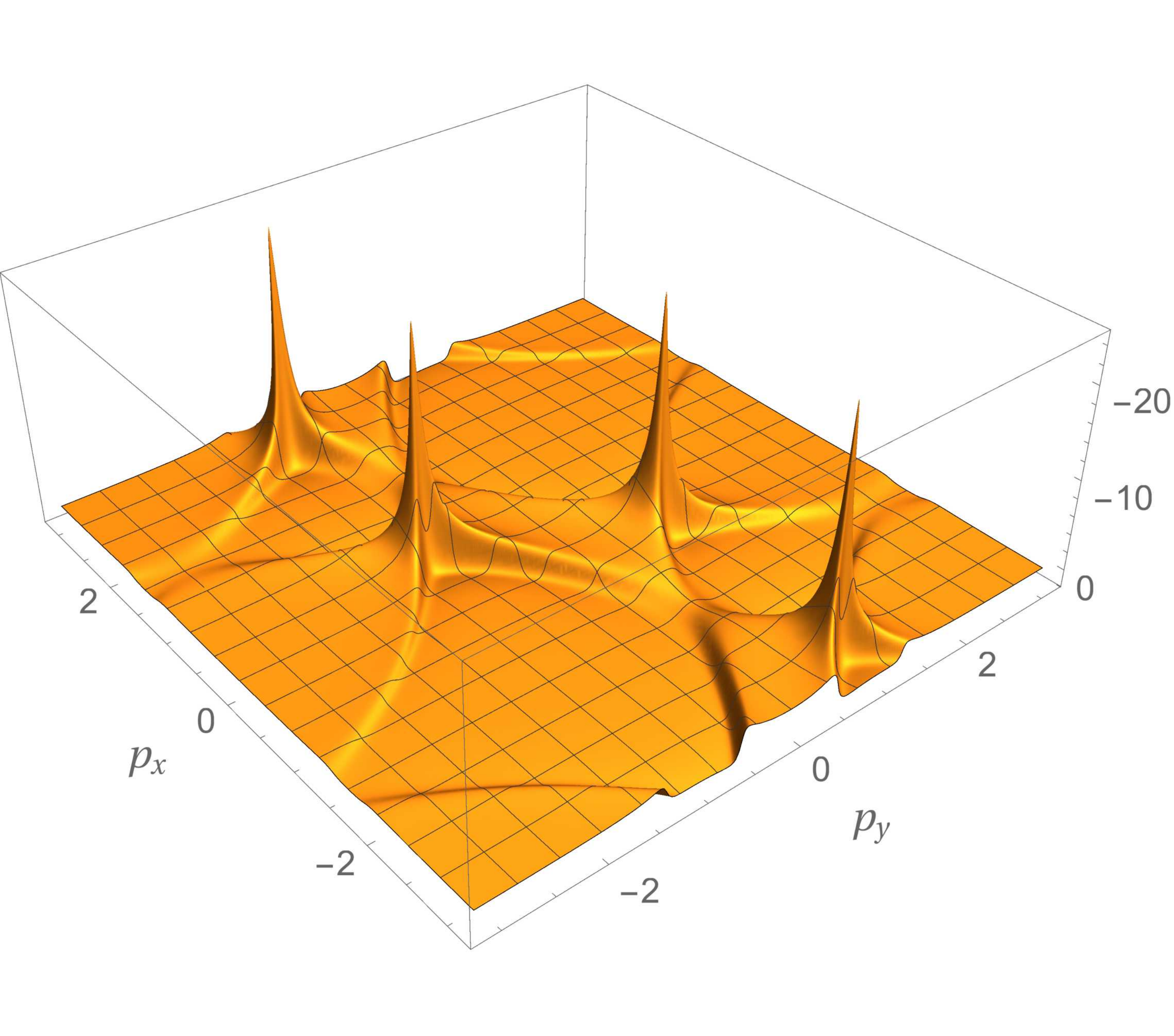}
    \caption{\it The value of the integrand is plotted against the two-dimensional momentum 
        $\mathbf{p}$ for $\Omega=1.0$ (left plot) and $\Omega=0.1$ (right plot). In this 
        example case the external momentum $\mathbf{l}$ is set to $(3.14,\, 0.78)$ and both form-factor 
        indices label the lowest order function, which is constant in momentum space.}
    \label{fig:integrand}
\end{figure}

The calculation of the two-particle coupling $V$ using a direct 
implementation of Eqs.~\eqref{eqn:su2fl}--\eqref{eqn:phd-su2fl} is 
numerically challenging. Even if the dependence on the external 
frequencies $k_{0,1}$, $k_{0,2}$ and $k_{0,3}$ is neglected---as 
we will do in the following---the scaling of the number of differential 
equations with respect to the number of momentum sampling points is cubic.
Using frequency independent two-particle couplings, the frequency 
integrals from Eqs.~\eqref{eqn:pp-su2fl}--\eqref{eqn:phd-su2fl} involve
just the $\Omega$ derivative of a product of two
fermionic propagators and can be performed analytically. The result of this shows sharp structures as 
function of momentum at small values of $\Omega$ (see Fig.~\ref{fig:integrand}).
As mentioned above, at low energy scales the system can get close 
to a phase transition, which is indicated by a strong increase of 
specific components of $V$. Thus a product of two strongly peaked 
two-particle couplings and sharp structured propagators 
constitutes the integrands of the momentum integrals in 
Eqs.~\eqref{eqn:pp-su2fl}--\eqref{eqn:phd-su2fl}.

In order to change Eqs.~\eqref{eqn:su2fl}--\eqref{eqn:phd-su2fl} in the 
direction of a numerically easier treatment, accompanied by the 
introduction of quantities with a more direct physical 
interpretation, we perform modifications that can be classified in 
three steps.\footnote{See Ref.~\cite{Lichtenstein2016} for a
more detailed derivation and an example application of the scheme.} 
First, the initial part $V^{(0)}$ is separated from 
the two-particle coupling and the rest is split into three 
single-channel coupling functions $\Phi^\mathrm{P}$, $\Phi^\mathrm{C}$ 
and $\Phi^\mathrm{D}$.
Their derivatives with respect to $\Omega$ are given by
$\mathcal{T}_\mathrm{pp}$, $\mathcal{T}^\mathrm{cr}_\mathrm{ph}$ and
$\mathcal{T}^\mathrm{d}_\mathrm{ph}$ respectively. This is motivated
by the fact that $V$ develops strong dependencies on the external
momentum combinations appearing in
Eqs.~\eqref{eqn:pp-su2fl}--\eqref{eqn:phd-su2fl} respectively, denoted by
$\mathbf{l}$ in the following.
In a second step, the remaining weak momentum dependencies of each
channel are expanded in a complete set of orthonormal functions
$\{f_n\}$---so-called form-factors. Since we can only use a finite
number of basis functions while doing numerics, we restrict the basis
to slowly oscillating functions to achieve a good description of weak
momentum dependencies of the channels. This latter step can be
interpreted as a sort of discretization with
$\Phi^\mathrm{P}_{\mathbf{l},\mathbf{k},\mathbf{k'}} \to
P_{m,n}(\mathbf{l})$---and equivalently for the $C$ and $D$ channels---%
where $\mathbf{k}$ and $\mathbf{k'}$ are replaced by form-factor
indices $m$ and $n$. As a consequence of implementing the first two
steps in Eqs.~\eqref{eqn:su2fl}--\eqref{eqn:phd-su2fl}, the scaling of the
number of coupled differential equations respect to the number of
momentum sampling points is reduced to a linear relation. Moreover,
the scaling respect to the number of form-factors is less important in
most cases, since a good description can be achieved even using a
small number of form-factors.

In a third and final step we change the form of the RHS of the
resulting differential equations by inserting two partitions of unity
of the form-factor basis set.
The fermionic propagators can then be separated from the two-particle coupling 
terms and the differential equation~(\ref{eqn:su2fl}) now takes the form
of three separate equations
\begin{align} 
    \label{eqn:matmul-P}
    \dot{\mathbf{P}} (\mathbf{l}) &= \mathbf{V}^P (\mathbf{l})\, \dot{\boldsymbol\Chi}^\mathrm{pp} (\mathbf{l}) \, \mathbf{V}^P (\mathbf{l})\, ,\\ 
    \label{eqn:matmul-C}
    \dot{\mathbf{C}} (\mathbf{l}) &= - \mathbf{V}^C (\mathbf{l})\, \dot{\boldsymbol\Chi}^\mathrm{ph} (\mathbf{l}) \, \mathbf{V}^C (\mathbf{l})\, ,\\ 
    \label{eqn:matmul-D}
    \dot{\mathbf{D}} (\mathbf{l}) &=  2  \mathbf{V}^D (\mathbf{l})\, \dot{\boldsymbol\Chi}^\mathrm{ph} (\mathbf{l}) \, \mathbf{V}^D (\mathbf{l}) -
   \mathbf{V}^C (\mathbf{l})\, \dot{\boldsymbol\Chi}^\mathrm{ph} (\mathbf{l}) \, \mathbf{V}^D (\mathbf{l}) - \mathbf{V}^D (\mathbf{l})\, \dot{\boldsymbol\Chi}^\mathrm{ph} (\mathbf{l}) \, \mathbf{V}^C (\mathbf{l}) \, ,
\end{align}
where 
\begin{align}
 \label{eqn:chi_pp}
 \Chi^\mathrm{pp}_{m,n} (\mathbf{l}) &= \int \! d\mathbf{p} \, \left[\int \! dp_0 \, G \left( p_0, \frac{\mathbf{l}}{2} + \mathbf{p} \right) \, G \left( -p_0, \frac{\mathbf{l}}{2} - \mathbf{p} \right)\right] \, f_m (\mathbf{p}) \, f_n (\mathbf{p}) \, ,\\
 \label{eqn:chi_ph}
 \Chi^\mathrm{ph}_{m,n} (\mathbf{l}) &= \int \! d\mathbf{p} \, \left[\int \! dp_0 \, G \left( p_0, \mathbf{p} + \frac{\mathbf{l}}{2} \right) \, G \left( p_0, \mathbf{p} - \frac{\mathbf{l}}{2} \right)\right] \, f_m (\mathbf{p}) \, f_n (\mathbf{p}) \, .
\end{align}
$\mathbf{V}^P$, $\mathbf{V}^C$ and $\mathbf{V}^D$ are two-particle
couplings with two momenta replaced by form-factor indices, and can be computed from $P$, $C$, and $D$ in the aforementioned inter-channel projections.
The inserted partitions of unity are also truncated by ignoring strongly
oscillating form-factors. Inner integrals from Eqs.~(\ref{eqn:chi_pp})
and (\ref{eqn:chi_ph}) can be treated analytically, while the
calculation of the outer (momentum) integrals requires a sophisticated
numerical integration scheme. Due to the last modification, we call the
scheme described in Eqs.~\eqref{eqn:matmul-P}--\eqref{eqn:chi_ph}
\textit{Truncated Unity fRG}
(TUfRG). 
\newcommand{\ml}{\mathbf{l}}

Numerically, this scheme is implemented in four steps organized in a
loop mimicking the flow of the ODE for decreasing values of
$\Omega$. Within the loop, the most intensive part of the computation
is given by the numerical integration. In the current C++
implementation of TUfRG, the numerical integration is parallelized using the
MPI+OpenMP paradigm. Each MPI process receives a subset of values of
$\ml$ indices while an OpenMP {\tt parallel for} pragma encapsulates
the actual computation of the integrals for all $m$ and $n$
values. Each integral is then assigned to a thread and computed
sequentially using the DCUHRE library~\cite{Berntsen1991}.
\begin{equation*}
  \label{eq:cycle}
	\begin{array}[l]{ccc}
          \color{orange}{\fbox{\rule[-0.2cm]{0cm}{0.6cm} \shortstack{{\small
          Assemble interaction}\\$P$,$C$,$D\,\rightarrow\,V^P$\\
          {\small\bf $\sim 20$\% CPU Time}}}} &
          \longrightarrow & \color{red}{\fbox{\rule[-0.2cm]{0cm}{0.6cm}
          \shortstack{{\small Perform 2D integration}\\
          $\dot\chi^\mathrm{pp}_{m,n} (\ml) \quad\ \forall\ m,\!n,\!\ml$\\
          {\small\bf $\sim 80$\% CPU Time}}}} \\[4mm]
          \uparrow & & \downarrow  \\[2mm]
          \color{blue}{\fbox{\rule[-0.2cm]{0cm}{0.6cm} \shortstack{{\small 
          Iterate ODE for $P$, $C$, $D$}\\$\frac{d}{d\Omega} P_{m,n}(\ml)$\\
          {\small\bf < 1\% CPU Time}}}} &
          \longleftarrow & \color{blue}{\fbox{\rule[-0.2cm]{0cm}{0.6cm} 
          \shortstack{\small Matrix multiplication \\ {\scriptsize$\sum_{p,q} 
          V^P_{m,p}(\ml)\, \dot{\chi}^\mathrm{pp}_{p,q}(\ml) \, V^P_{q,n}(\ml)$}\\
          {\small\bf < 1\% CPU Time}}}}
 	\end{array}
\end{equation*}
Using example values of $\mathbf{l}$, $m$ and $n$,
Fig.~\ref{fig:integrand} shows the integrand from
Eq.~(\ref{eqn:chi_pp}) at a high and a low value of $\Omega$. This
example illustrates a general characteristic of the integrands: While
at high $\Omega$ values the variations in momentum space are smooth,
sharp edges and peaks emerge as the energy scale is lowered.  This
means in terms of numerical integration that the density of sampling
points in momentum space should be chosen adaptively and separately
for every integration. As the data from Fig.~\ref{fig:integrand}
suggest, the adaptive routine used should furthermore be able to
refine the grid of sampling points using strongly local criteria in
order to reduce the inaccuracies caused by sharp structures and to
save time when flat regions are considered. In the next section we
show how such a target is achieved by Algorithm \ref{adapt-par}, 
and give an account of its parallel implementation.


\section{Adaptive integration a-la Clenshaw-Curtis}
\label{sec:parintad}

The main target of adaptive integration is to decrease the error in a
consistent and controlled fashion over relatively low-dimensional
domains~\cite{Cools:2002uc}. A possible choice to increase accuracy is
to increase the number of integration points for a given integration
method. Alternatively one can fix the number of integration points and
instead partition the integration domain. Numerical
integration with the same number of integration points is then
performed on each subdomain. The latter method is known as
\textit{adaptive integration}.  One of the most accurate variations of
such a method computes the error across the whole domain\cite[Chapter
6]{Krommer:1994kz}. If some
estimate for the global error is above a given threshold, one
iteratively subdivides and integrates the sub-domain with the largest
local error.  Algorithm \ref{globerr} shows the typical structure of such
an adaptive integration scheme.
\begin{algorithm}[h!t]
	\caption{Adaptive integration on domain $\mathcal{D}$ with global error criterion}
	\label{globerr}
	\begin{algorithmic}
		\Function{adaptive}{Integrand $\phi,$ Domain $\mathcal{D},$ Target Error $\varepsilon$}
                       \State Compute integrals 
                       \Call{$\operatorname{Q}$}{$\phi,\mathcal{D},n_1$}
                       and \Call{$\operatorname{Q}$}{$\phi,\mathcal{D},n_2$}
			\State $\err \phi = \lvert
                       \operatorname{Q}_{n_1}\phi - \operatorname{Q}_{n_2}\phi \rvert$
			\State Store domain $\mathcal{D}$ and $\err \phi$
			\While {$ \err \phi > \varepsilon $}
				\State Take the sub-domain $\mathcal{D}^s$ with largest error
				\State Subdivide it into parts
                               $\mathcal{D}^{s_0},\ldots,\mathcal{D}^{s_{d-1}}$
                               \For{$a=1:d$}
                                        \State Compute \Call{$\operatorname{Q}$}{$\phi,\mathcal{D}^{s_a},n_1$}
                                        and \Call{$\operatorname{Q}$}{$\phi,\mathcal{D}^{s_a},n_2$}
                                        \State $\textsc{err}^{s_a}[\phi] = \lvert
                                        \operatorname{Q}^{s_a}_{n_1}\phi
                                        - \operatorname{Q}^{s_a}_{n_2}\phi \rvert$
			                 \State Store domain
                                         $\mathcal{D}^{s_a}$ and $\textsc{err}^{s_a}[\phi]$
                               \EndFor
				\State $\err{\phi} = \displaystyle{\sum_{s}}\ \textsc{err}^{s}[\phi]$
			\EndWhile
			\State \Return {\sc value} = $\displaystyle{\sum_{s}} 
                        \operatorname{Q}\left(\phi,\mathcal{D}^{s},n_2\right)$
		\EndFunction
	\end{algorithmic}
\end{algorithm}

In the current work we compute the numerical integrals (and also the
estimate for the error) using the Clenshaw-Curtis Quadrature (CCQ)
formula\footnote{For a review of Clenshaw-Curtis and a comparison with
  Gauss quadrature rules we refer to the excellent review
  \cite{Trefethen:2008hw}}. Starting with $n_1$ integration points, we
settle for a formula with $n_2 \geq n_1$ as the more accurate
estimate\cite{BERNTSEN:1989wk}. With this choice the error estimate is equal to
\[
\err \phi = \lvert \operatorname{Q}_{n_1} \phi - \operatorname{Q}_{n_2} \phi
       \rvert ,
\]
where with
$\operatorname{Q}_{n} \phi = \operatorname{Q} \left(
  \phi,\mathcal{D},n\right)$
we indicate the computation of the integral $\Phi = \int_{\mathcal{D}} \phi$ over the domain
$\mathcal{D}$ through numerical quadrature with $n$ integration
points\footnote{We use the conventional notation indicating with
  the capital symbol the integral ($\Phi$) and with the corresponding small cap
  symbol ($\phi$) its integrand.}. When $n_2$ is proportional to $n_1$, the advantage of the CCQ
scheme---compared to Gauss for instance---is the reuse of the $n_1$
points as a subset of the $n_2$ points. In the rest of this work we
set $n_1 = N$ and $n_2 = 2 n_1$.

The schematic description in Algorithm \ref{globerr} should be applied
to the computations of the integral on the RHS of the flow
Eqs.~\eqref{eqn:pp-su2fl}--\eqref{eqn:phd-su2fl}. After discretizing
and  projecting  (using a truncated partition of unity), the RHS of
such equations are split in multiplications of two-particle couplings
$\mathbf{V}^i \ (i=P,C,D)$ and susceptibility factors
${\boldsymbol\Chi}^{j}$ ($j={\rm pp,ph}$), only the latter expressed in
terms of integrals. Despite the fact that now the integrals seem
limited to the RHS of Eqs.~\eqref{eqn:chi_pp}--\eqref{eqn:chi_ph}, the
adaptive approach has to encompass the whole set of integrals labeled
by the indices $m$ and $n$.  Moreover, the original integrals included
also the values of the couplings $\mathbf{V}^i$ in the integrand, so
these quantities play an active role in the computation of the global
error. Let us clarify this point by considering just the
particle-particle channel Eq.~\eqref{eqn:pp-su2fl}, and formally
evaluating it using a generic quadrature formula
\begin{equation}
\operatorname{Q}_{N}\tau_\mathrm{pp} = - \sum_{\ell=1}^{N} w_\ell
\left[ \partial_\Omega G(p_\ell) \, G(k_1+k_2-p_\ell) \right]
  V (k_1,k_2,p_\ell) \, V(k_1+k_2-p_\ell,p_\ell,k_3) \, ,
\end{equation}
where the $w_\ell$ are the weights associated with the quadrature points
$p_\ell$. After some rearrangements and the introduction of the truncated
partition of unity, the RHS of this equation is transformed into 
\(
\sum_{p,q} V_{m,p}^P\, 
\left(\operatorname{Q}_{N}\dot{\chi}_{p,q}^{\mathrm{pp}} 
\right) \, V_{q,n}^P\,
\)
where we made explicit the $m$ and $n$ indices and suppressed, for the
moment, the dependence on the $\mathbf{l}$ index.  Despite the fact
that now this quantity is the sum of distinct quadratures
$\operatorname{Q}_{N}\dot{\chi}_{p,q}^{\mathrm{pp}}$, the global error
should be thought as defined by the original expression
$\left|\operatorname{Q}_{N}\tau_\mathrm{pp} -
\operatorname{Q}_{2N}\tau_\mathrm{pp}\right|$, leading to the following expression
\begin{equation}
\label{eq:P_globerr}
\begin{split}
\err{\dot{P}_{m,n}} &  = \left|\sum_{p,q} \left[ V_{m,p}^P\, 
\left(\operatorname{Q}_{N}\dot{\chi}_{p,q}^{\mathrm{pp}}\right) 
V_{q,n}^P - V_{m,p}^P\, \left(\operatorname{Q}_{2N}\dot{\chi}_{p,q}^{\mathrm{pp}} 
 \right) V_{q,n}^P \right]\right| \\
 & \leq \left\| V_{m,:}^P\, \right\|_\infty
   \left\| V_{:,n}^P \right\|_\infty
   \sum_{p,q}\left|\operatorname{Q}_{N}\dot{\chi}_{p,q}^{\mathrm{pp}} 
   - \operatorname{Q}_{2N}\dot{\chi}_{p,q}^{\mathrm{pp}} \right| ,
\end{split}
\end{equation}
where $x =V_{m,:}$ is the vector made by all column entries
corresponding to the $m^{th}$ row and $\|x\|_\infty = \max_j
|x_j|$.
\begin{algorithm}[t!h]
\caption{Parallel adaptive integration of TUfRG with global error}
\label{adapt-par}
\begin{algorithmic}[1]
\ForAll{$i,j,k,\mathbf{l}$}
    \Function{adaptive}{Integrand $\phi^{i,j,k},$ Domain $\mathcal{D},$ Target Error $\varepsilon$}
          \State done \(=\) \textit{false}
           \While { done \(\neq\) \textit{true} } \label{ln:bgnWhile}
                  \State Take the domain $\mathcal{D}^{s} \subseteq
                  \mathcal D$ and indices $(p,q)$ with largest error \label{ln:load}
                  \State Subdivide it into parts
                  $\mathcal{D}^{s_0},\ldots,\mathcal{D}^{s_3}$
                  \For{$a=1:4$}
                           \State Compute \Call{$\operatorname{Q}$}{$\chi_{p,q}^j,\mathcal{D}^{s_a},N$}
                           and \Call{$\operatorname{Q}$}{$\chi_{p,q}^j,\mathcal{D}^{s_a},2N$}
                           \State $\textsc{err}^{s_a}[\chi_{p,q}^j] = \lvert
                           \operatorname{Q}^{s_a}_{N}\chi_{p,q}^j 
                           - \operatorname{Q}^{s_a}_{2N}\chi_{p,q}^j \rvert$
                           \State Store domain $\mathcal{D}^{s_a}$,
                           indices $(p,q)$ and
                           $\textsc{err}^{s_a}[\chi_{p,q}^j]$ 
                           \label{ln:store2}
                  \EndFor
                  \State $\err{\phi^{i,j,k}} =
                  \displaystyle{ \sum_{s,p,q}} 
                  \textsc{err}^{s}[\chi_{p,q}^j]$
                  \IIf { \(\err{\phi^{i,j,k}} < \varepsilon \) } done \(=\) \textit{true}
                  \EndIIf
          \EndWhile \label{ln:endWhile}
          \State \Return $\left(\textsc{value}\right)^j_{p,q} = \displaystyle{
            \sum_{s}} \operatorname{Q}\left(\chi_{p,q}^j,\mathcal{D}^{s},2N\right)$
          \EndFunction
   \EndFor
\end{algorithmic}
\end{algorithm}

We can think of
the entire numerical integration as the union of the quadratures
$\operatorname{Q}_{n}\dot{\chi}_{p,q}^{\mathrm{pp}}$ on the same
domain $\mathcal{D}$ for each value of the indices $p$ and $q$.  While
each adaptive quadrature labeled by $p$ and $q$ returns its own {\sc
  value}, the absolute error is computed globally over all indices
$(p,q)$. We further simplify the definition of the error by dropping
the terms proportional to $\left\| V \right\|_\infty$. This last step may
seem arbitrary but it is in part justified by the fact that, in the
actual computation, we are only interested in  the error relative to the
value of the function. In order to maintain generality we define with
$\phi^{i,j,k} = \mathbf{V}^i \dot{\boldsymbol\chi}^j \mathbf{V}^k$ and
the associated global relative error as
\[
\err{\phi^{i,j,k}} = 
\sum_{p,q} \err{\dot{\chi}_{p,q}^j}. 
\] 
We kept the index $\mathbf{l}$ still implicit so as to avoid cluttering
the notation, but it is understood that all definitions above depend
implicitly on it. With these definitions in mind we end up with the
adaptive quadrature scheme illustrated in Algorithm \ref{adapt-par}.

\section{Parallel implementation}
\label{sec:impl}

In this work, we describe a parallel implementation of the {\sc adaptive}
function of Algorithm \ref{adapt-par} over one computing node using OpenMP
pragmas, and leave the outer {\bf for} loops---identified by indices $i$, $j$,
$k$, and $\mathbf{l}$---distributed over MPI processes.  Each elementary
integration is encoded as a {\tt task}, which can be imagined as a struct type.
Each {\tt task} has the following members: an {\tt id} field that corresponds to
distinct values of the $p$ and $q$ indices, a {\tt domain}, the two values {\tt
  val\tus N} and {\tt val\tus 2N} computed according to the CCQ method, and
an estimate of the error {\tt err}.

The adaptive integration scheme requires the tasks with the largest error to be
scheduled first.  Such an approach is not easily expressible with the OpenMP
task construct.  Although OpenMP tasks have recently gained support for task
priorities, the allowed priority values are limited to non-negative scalars.  As a
result PAID cannot make use of OpenMP tasks.

The container into which the tasks are placed is a heap data structure that uses
the {\tt err} as the sorting key. A heap structure allows cheap en- and
dequeuing of tasks.  The heap is initialized at the beginning of the program.
\begin{lstlisting}[
    caption={Initialization of the task queue container},
    label=alg:init
]
    $\err{\phi} \coloneqq 0.0$
    for all $(p,q)$
        !{\tt task.id}! $\coloneqq (p,q)$ and !{\tt task.domain}! $\coloneqq \mathcal{D}$
        !{\tt task.val\tus N}! $\coloneqq \operatorname{Q}_{N}\chi$ and !{\tt task.val\tus 2N}! $\coloneqq \operatorname{Q}_{2N}\chi$ 
        !{\tt task.err}! $ \coloneqq \lvert \operatorname{Q}_{N}\chi - \operatorname{Q}_{2N}\chi\rvert$
        !{\tt container}.{\sc push}({\tt task})!
        $\err{\phi} \pluseqq$ !{\tt task.err}!
    !{\sc heapify}({\tt container}, key = {\tt task.err})!
\end{lstlisting}

The heap structure of the container guarantees that task extraction
is done in a way that refines regions with larger error
estimates first, independent of which pair of indices $(p,q)$ they
belong to. In a way this algorithm can be seen as adaptive integration
with starting regions defined by both \(\mathcal{D}\)
and \((p,q)\).
Due to the adaptivity based on the global error, the OpenMP parallel block has to
enclose the domain $\mathcal D$ as well as the indices $(p,q)$. As previously
stated, PAID cannot make use of OpenMP's more advanced work sharing constructs.
Instead, Algorithm \ref{adapt-par} parallelizes the main part of the routine from
line \ref{ln:bgnWhile} to \ref{ln:endWhile} using just the \verb+parallel+ directive
(see line \ref{ln:ompar} of Listing \ref{alg:final}). Access to the queue
in lines \ref{ln:load} and \ref{ln:store2} requires exclusive access in order to
avoid race conditions.  For queues that are not thread-safe a mutex is required
(\verb+critical+ directive). This may decrease parallel performance as threads
may need to wait for access to the queue. 
For this reason we implement bulk extraction and insertion into a thread-local container:
Each thread can extract a MaxTask number of tasks,
whose value is set by the user. 
Care has to be taken in choosing MaxTask; Its optimal value is a
trade-off between maintaining acceptable levels of parallel
performance and avoiding unnecessary adaptive refinements.
\begin{lstlisting}[
    caption={Extract tasks with maximal error from the queue},
    label=alg:crit1
]
    !\#{\sc pragma omp critical} \{!
    for $n = 1 : \text{MaxTask}$
        !{\tt local\_container}$[n] =$\ {\sc extract-max}({\tt container})! 
    !\}!
\end{lstlisting}

This results in a work sharing construct, as each task returned from the heap is different.
Tasks are processed by partitioning their domain once in each dimension, which yields four new tasks.
Before the new tasks can be inserted into the heap an error estimate is required, which in turn requires evaluation of the integrals.
\begin{lstlisting}[
    caption={Divide domains and evaluate new tasks},
    label=alg:funceval
]
    for $n = 1:\text{MaxTask}$
        !{\tt evaltask}$[n,1:4]$! $\coloneqq$ Divide !{\tt local\_container}$[n]$.{\tt domain}! into 4 parts
        for $a = 1:4$
            !{\tt evaltask}$[n,a]$.{\tt domain}! $\coloneqq$ part $a$ of !{\tt local\_container}$[n]$.{\tt domain}! 
            Compute !{\tt evaltask}$[n,a]$.{\tt val\tus N} and {\tt evaltask}$[n,a]$.{\tt val\tus 2N}!
            Compute !{\tt evaltask}$[n,a]$.{\tt err}!
\end{lstlisting}
Eventually, the global error is updated within the mutex. Each thread then inserts the new tasks,
together with their relative sub-domain, and {\tt id} in the heap.
\begin{lstlisting}[
    caption={Update error and insert new tasks in the queue},
    label=alg:crit2
]
    !\#{\sc pragma omp critical} \{!
    for $n =1:\text{MaxTask}$
        !$\err{\phi} \minuseqq$ {\tt local\_container}$[n]$.{\tt err}!  
        for $a = 1:4$
           !$\err{\phi} \pluseqq$ {\tt evaltask}$[n,a]$.{\tt err}!
           !{\sc insert}({\tt evaltask}!$[n,a] \Rightarrow$ !{\tt container})! 
    !\}!
\end{lstlisting}
%
%
%
The termination criterion need only be checked by a single thread at the end of its
block of refinements. This implies that the termination criterion is
checked not sooner than after MaxTask refinements. When the global
error is lower than the required threshold, all other threads are
instructed to exit the while loop via the \verb+done+ flag. The entire
program, which includes all previous Listings, is illustrated in Listing
\ref{alg:final}.
\begin{lstlisting}[
    caption={Full program},
    label=alg:final,
]
    Program !{\sc PAID}!($\phi,\mathcal{D},\varepsilon$)
       done !\(\coloneqq\) \textit{false}!
       Initialize the task queue !{\tt container} \hfill (Listing~\ref{alg:init})!
       !\#{\sc pragma omp parallel} \{ \label{ln:ompar}!
       while done !\(\neq\) \textit{true}! do
             Extract tasks with max error from !{\tt container} \hfill (Listing~\ref{alg:crit1})!
             Divide the domain and evaluate new tasks !\hfill (Listing~\ref{alg:funceval})!
             Update $\err{\phi}$ and insert new tasks in the queue !\hfill (Listing~\ref{alg:crit2})!
             !\#{\sc pragma omp master} \{!
             if !\(\err{\phi} < \varepsilon \)! then done !\(\coloneqq\) \textit{true}!
             !\}!
       !\}!
       !\textbf{forall} distinct {\tt task.id} $=(p,q)$!
           return !$\left({\sc value}\right)_{(p,q)} \coloneqq \sum_{{\tt task.domain}}${\tt task.val\tus 2N}!
\end{lstlisting}


\section{Results and conclusions}
\label{sec:rescon}

In order to illustrate the effectiveness of PAID within the TUfRG code, we
present a number of numerical tests, run on the JURECA computing cluster located
at the J\"ulich Supercomputing Centre. Each node of the cluster is equipped with
2 Intel Xeon E5-2680 v3 Haswell CPUs. All tests were run with a single MPI
rank per compute node.  Node level parallelism is exclusively due to the shared
memory parallelization of the adaptive quadrature implementation described in
Algorithm \ref{adapt-par}.
%

In the following we draw a comparison between the previous implementation 
using DCUHRE and the newly developed implementation based on PAID.
As both adaptivity and parallel efficiency play an important role 
in terms of performance, we conducted the comparative analysis in 
terms of these two aspects separately, before we compare the runtimes. 

\begin{figure}[t]
    \centering
    \includegraphics[width=0.45\textwidth]{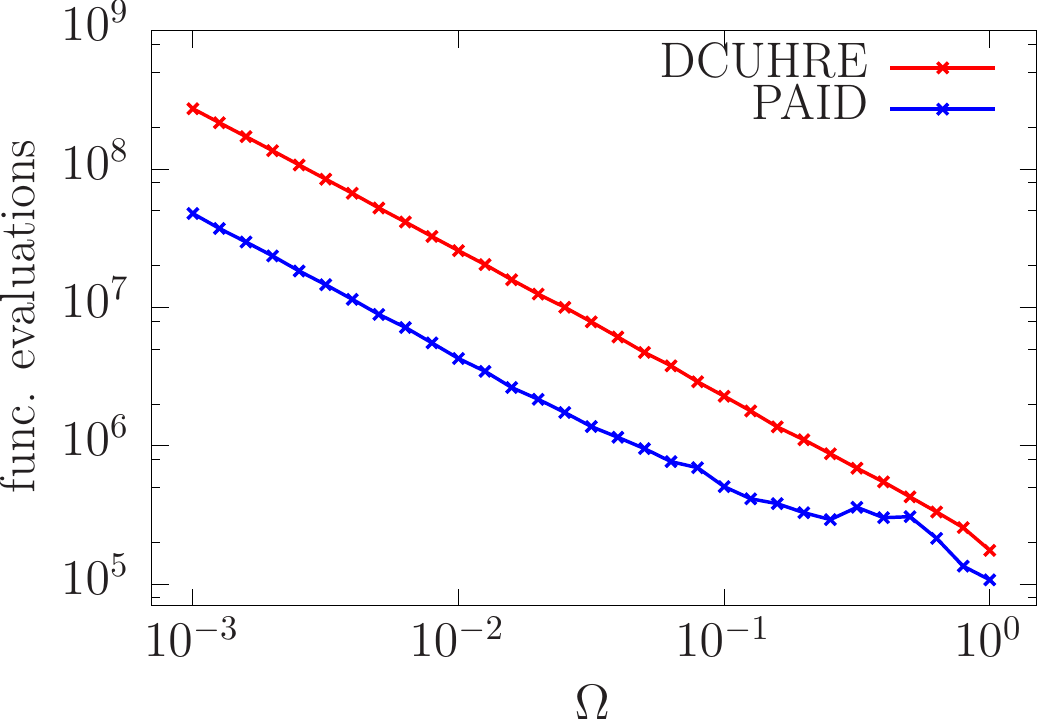}
    \hspace{0.08\textwidth}
    \includegraphics[width=0.45\textwidth]{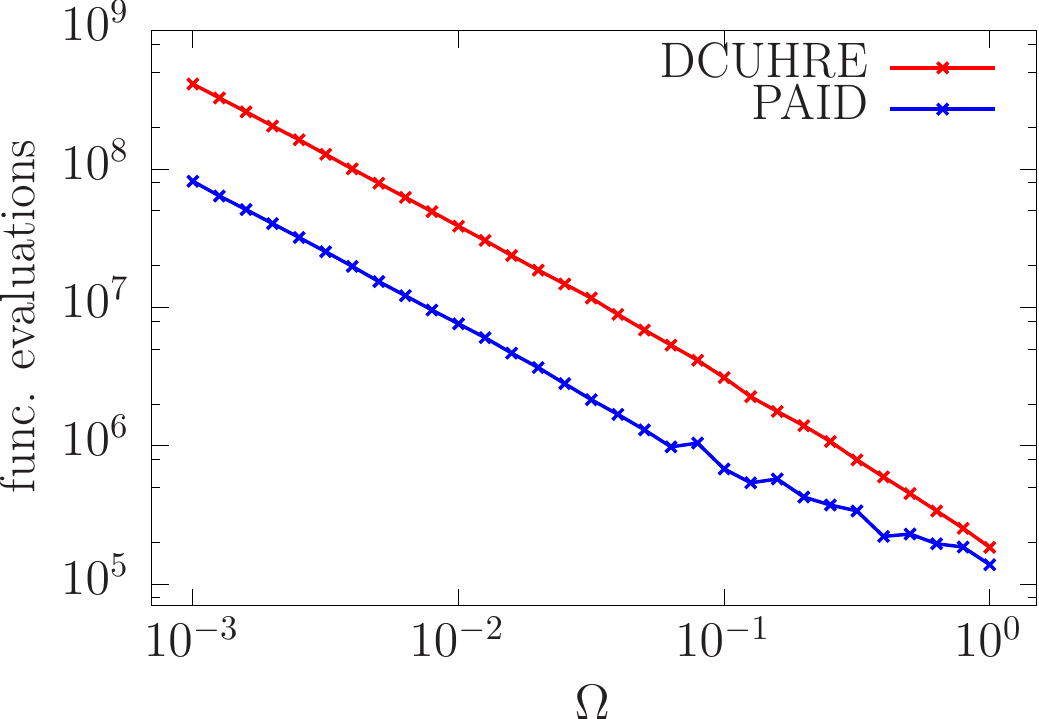}
    \caption{\it The number of function evaluations needed for  
        calculating all integrals of a fixed $\mathbf{l}$ value 
        is plotted against $\Omega$. We use a form-factor expansion
        that results in $45$ independent integrals for each external 
        momentum $\mathbf{l}$, which is fixed to $(1.57,1.31)$ 
        (left plot) and $(2.88,0.26)$ (right plot) respectively. 
        The results of both implementations---the one using 
        DCUHRE (red) and the one using PAID (blue)---are shown in 
        the same plot in favor of a direct comparison. In order 
        to use the same number of evaluations per subregion as in 
        DCUHRE, we set the PAID parameter $N$ to $4$. 
        Further we use MaxTask$\ =10$.}
    \label{fig:nevals}
\end{figure}
Fig.~\ref{fig:nevals} shows that the number of function evaluations
needed by PAID is smaller than the one needed by DCUHRE at all values
of the scale parameter, especially at low scales where most of the
computation time is used.%
\footnote{Notice that the fRG flow in the current setup starts at high
  $\Omega$ values and successively reduces this scale during the
  flow.}  As a smaller number of evaluations implies a more efficient
partition of the integration domain, this number can be seen as an
inverse measure for the adaptivity of the implementation provided by
PAID, especially at low scales where the integrands tend to blow
up. The difference in adaptivity between the both schemes can be
understood as a consequence of the error estimation. DCUHRE computes
the errors of the integrals labeled by $p$ and $q$ in isolation so
that each error fulfills the same termination criterion
independently. The PAID scheme treats all integrals as \textit{one}
task which in practice prioritizes computation over the most difficult
partitions of the domain. As the major part of the computation time is
used by the low-$\Omega$ integration domains, a code that is more
efficient in this region of the parameter space pays off in terms of
total runtime.

\begin{figure}[t]
    \centering
    \includegraphics[width=0.45\textwidth]{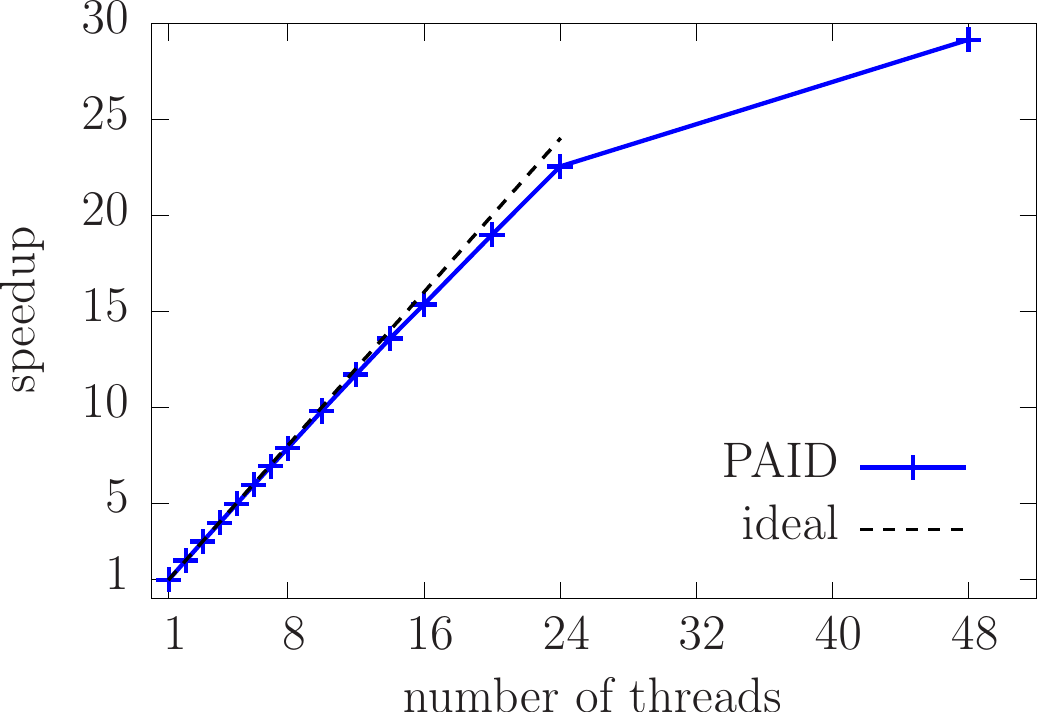}
    \hspace{0.08\textwidth}
    \includegraphics[width=0.45\textwidth]{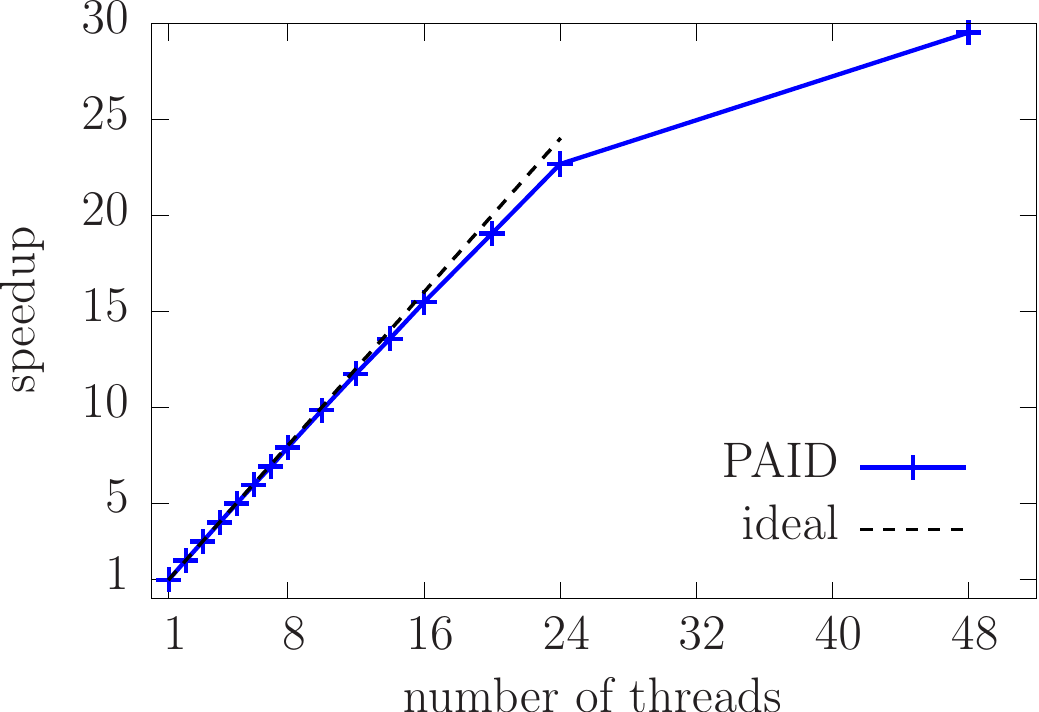}
    \caption{\it These plots show the speedup at $\Omega=10^{-3}$ against 
        the number of threads
        for the implementation based on PAID. For thread numbers up to $24$ 
        each compute core executes only a single thread. At $48$ threads
        each compute core processes two threads at a time using
        simultaneous multithreading. The $\mathbf{l}$-values are 
        chosen as in Fig.~\ref{fig:nevals} and there are 
        $325$ integrals to calculate per $\mathbf{l}$. For this analysis we 
        use the PAID parameters that result in the best performance: $N=6$ 
        and MaxTask$\ =18$.}
    \label{fig:speedup}
\end{figure}
The second performance analysis addresses the parallel efficiency of
PAID. Fig.~\ref{fig:speedup} shows that the speedup is close to ideal
for any thread number up to $24$\footnote{The granularity of the
  affinity is set to 'compact,core,1'.}.  Using SMT---up to $48$
threads---still increases the speedup compared to the one using $24$,
but the curves in Fig.~\ref{fig:speedup} show a slower increase in
performance. This result suggests that the code is
  compute bound and can not profit highly from a larger memory
  bandwidth per core. Although the shared memory parallelization of
the implementation of TUfRG using DCUHRE is much simpler---as
described in section~\ref{sec:tufrg}---,we find a speedup which is as
high as the one we achieve using PAID. We verified that the runtimes
of the integrals for distinct $p$ and $q$ within a fixed value of
$\mathbf{l}$ do not vary much in serial execution. In such a case the
parallelization over the $p$ and $q$ values does not suffer from load
imbalances and results in a close to ideal speedup.

\begin{figure}[t]
    \centering
    \includegraphics[width=0.45\textwidth]{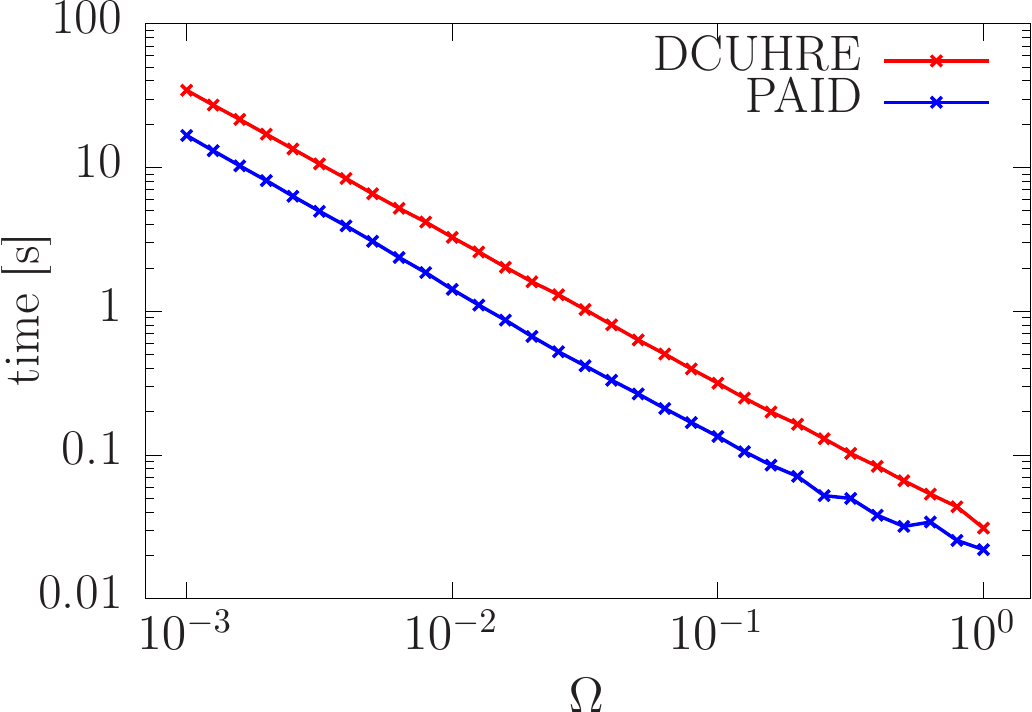}
    \hspace{0.08\textwidth}
    \includegraphics[width=0.45\textwidth]{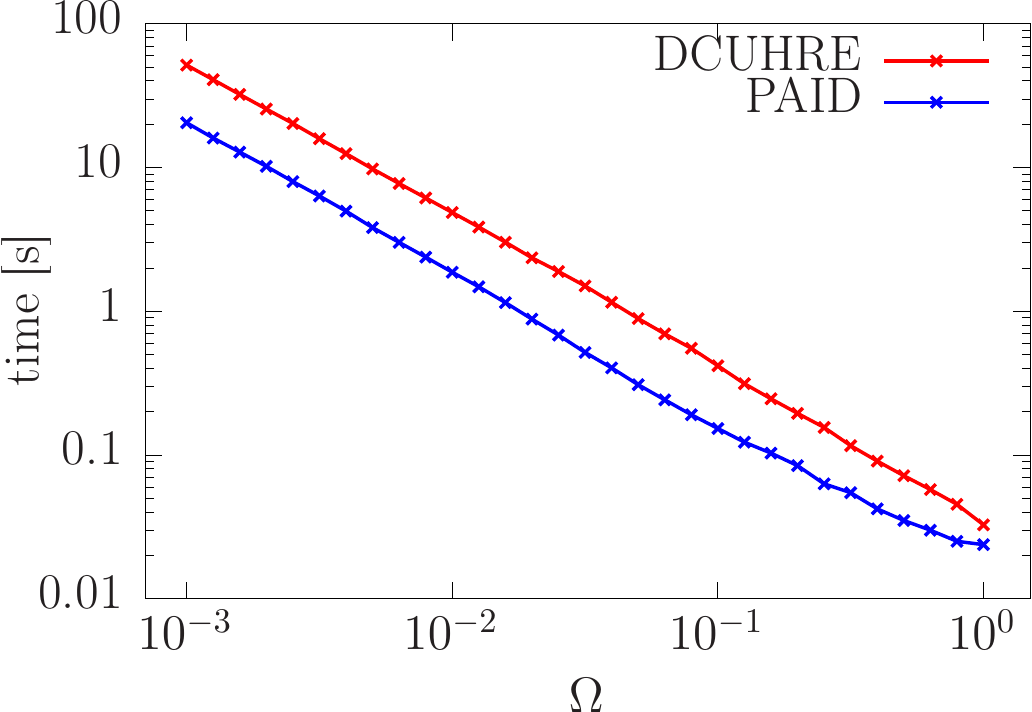}
    \caption{\it The computation time needed for calculating all integrals 
        of a fixed $\mathbf{l}$ value using $48$ threads is plotted 
        against $\Omega$. We use a 
        form-factor expansion that results in $325$ independent integrals 
        for each $\mathbf{l}$, which takes the same values as in 
        Fig.~\ref{fig:nevals}. For reasons of comparison we use $N=4$ in 
        PAID as in the analysis shown in Fig.~\ref{fig:nevals}. 
        Further, MaxTask is set to~$7$.}
    \label{fig:timings}
\end{figure}
In a third set of tests---possibly the most relevant to a user of
TUfRG---we compared the runtimes that are needed by DCUHRE and PAID to
perform all the integrations within a fixed value of $\mathbf{l}$. As
illustrated in Fig.~\ref{fig:timings}, PAID needs less compute time
than DCUHRE at all scales and is about $2$--$3$ times faster at low
$\Omega$ values.

In conclusion the TUfRG code greatly benefits from the proposed
adaptive integration algorithm both in terms of load balancing and
adaptivity. The result is a good exploitation of the node-level
parallelism at any stage of the flow equation without the need of
ad-hoc parameter choices. Comparing the new integration scheme with
the one provided by DCUHRE shows that the PAID algorithm exhibits a
higher level of adaptivity which in turn leads to shorter runtimes. In
addition, the use of standard OpenMP pragmas ensures performance
portability over clusters other than JURECA with the potential for
off-loading to many-cores platforms with minimal effort. In the
future, we envision to expand the internal parallelism of PAID to
distributed memory. Such an extension could replace the currently used
distribution of the $\mathbf{l}$ values over the MPI ranks and would
prevent load imbalances that limit the number of accessible nodes in
the current implementation.


\section*{Acknowledgements}
Financial support from the J\"ulich Aachen Research Alliance--High
Performance Computing, and the Deutsche Forschungsgemeinschaft (DFG)
through grants GSC 111, RTG 1995 and SPP 1459 is gratefully
acknowledged.  We are thankful to the J\"ulich Supercomputing Centre
(JSC) for the computing time made available to perform the numerical
tests. Special thanks to JSC Guest Student Programme which sponsored
the research internship of one of the authors.

\bibliography{NumInt}{}
\bibliographystyle{splncs}

\end{document}